# Tuning the structural, electronic and intrinsic transport properties of two-dimensional borophene sheets by strain

Vivekanand Shukla[1*], Anton Grigoriev[1], Naresh K. Jena[1*] and Rajeev Ahuja[1,2]

[1] Condensed Matter Theory Group, Materials Theory Division, Department of Physics and Astronomy, Uppsala University, Box 516, SE-75120, Uppsala Sweden

[2] Applied Materials Physics, Department of Materials and Engineering, Royal Institute of Technology (KTH), SE-10044, Stockholm Sweden

Corresponding Author: VS (vivekanand.shukla@physics.uu.se); NKJ (jenanaresh@gmail.com)

**ABSTRACT**

The success of graphene as a 2D material has opened new research paradigm and the long-standing quest for 2D analogues. Two recent reports of realization of borophene (Science, 350, 1513-1516 (2015); Nature Chemistry, 8, 563–568 (2016)) focus on the inherent anisotropy and directional dependent electronic properties in borophene polymorphs. We use first principle density functional theory (DFT) calculations to study structural, electronic and transport properties of two borophene polymorphs $\beta_{12}$ and $\chi$ borophene. To tune the electronic and transport properties of borophene, application of strain is a straightforward approach and we find that strain as low as 6% brings remarkable changes in the electronic and transport properties of these two structures. We verified the directional dependency in the electron transport properties in these two boron polymorphs and find tunable anisotropic behavior of the transport properties in these 2D materials. We further investigate current-voltage (IV) characteristics in low bias regime after applying strain on these systems to see how this external strain affects the anisotropy of current. These new materials have raised many open possibilities, and accomplishing stable 2D borophene structures may lead to some degree of strain in the system because of the substrate lattice mismatch. Our findings concerning sizeable tuning of transport and IV characteristics at the expense of minimal strain, suggest the suitability of 2D borophene for futuristic device applications.

## I. Introduction

Graphene's rise to prominence has steered research on two-dimensional (2D) materials and subsequent quest for other counterparts of graphene in different electronic characteristics regimes like semiconductor, metal, or insulator is actively pursued so as to make the dream of atomically thin device come true[1–5]. Monolayers of some novel 2D materials like graphene, transition metal dichlagonides (TMDC)[6–8], hexagonal boron nitride (h-BN)[9], monolayer black phosphorus (BP)[10] are quite accessible by robust synthetic methods and have been successfully demonstrated in nanoscale electronic devices[11–13]. In this quest, boron, the fifth element of the periodic table, showing tremendous strength and hardness along with possessing high melting point and being lightweight, has attracted considerable curiosity[14,15]. Boron can exist in several low dimensional polymorphs analogous to carbon in the form of fullerenes, clusters and nanotubes[16–19]. Low dimensional polymorphism in boron has been the subject for several insightful computational studies[20–22]. Synthesis of boron in 2D form has been a long standing challenge for the scientific community. Recently, very thin layer of boron (~0.8

Å), known as borophene, has been characterized on the silver surface under ultra-high vacuum conditions in line with a previously predicted theoretical structure, few years back[20,21,23]. There was another experimental effort to synthesize 2D boron, where two polymorphs were reported named as $\chi$ and $\beta_{12}$-borophenes[24–26]. These are planner in structure unlike to its previously reported counterpart (borophene), which has triangular buckled structure[23]. In the planar structure, triangular boron lattices with different arrangement of hexagonal hollow sites were reported. These two structures are stable against oxidation, which widens their scope for future potential applications[24]. It has been already predicted that the electronic structure of both the polymorphs are metallic[27]. These borophene structures have been reported as unstable in isolated form, but this instability could be overcome with vacancy defect and deposition on some non-reactive surface[28]. Existence of massless Dirac fermions in these two polymorphs makes them even more compelling towards the nano-electronic applications[29,30]. Additionally, these polymorphs have shown phonon meditated superconductivity[31], and because of the light weight, diverse electronic properties and mechanical stability, they have been reported as high capacity electrode for electrochemical applications[32–34] and as catalyst for hydrogen evolution reaction[35].

Typically, several approaches can be applied to engineer the electronic structure of 2D materials, like electrostatic gating and molecular adsorption, that have been used for graphene, phosphorene and others[36–39]. Moreover, electronic, transport and optical properties are considerably determined by atomic geometry, and it follows that strain is the most suitable way to tune intrinsic properties of these 2D materials. These experimentally synthesized polymorphs can be realized by change in substrate, growth conditions and substrate crystallographic orientation, which results in a diverse class of 2D metals. Furthermore, small strains, induced by the substrate, can be utilized to tune the properties of these polymorphs. This strain engineering has been broadly studied for 2D materials such as graphene, phosphorene and TMDCs. For example, strain in graphene reinforces the electron-phonon coupling[37,40,41]. Similarly, in $MoS_2$, small compressive(tensile) strain increase(decrease) the fluorescence intensity and further strain leads to semiconductor to metal transition[42,43]. Like the other counterparts, 2D borophenes can also withstand high strains without significant distortion in their structural properties.

In this work, using Density functional theory (DFT) calculations, we investigate the anisotropy in structural, electronic and transport properties of the two recently synthesized novel borophene ($\beta_{12}$ and $\chi$) polymorphs. We report that these two polymorphs can withstand unidirectional compressive and tensile strains up to 6% with the deformation energies less than 100 meV/atom. Interestingly, their electronic structures are significantly modified at these strains while they retain their metallic nature. Emphasizing the effect of external strain to tune the electronic and transport properties in $\chi$ and $\beta_{12}$-borophene, we put forward that retaining their metallic nature makes these 2D materials suitable for electrode application in nanoscale devices.

## II. Computational Details:

First principles density functional theory (DFT)[44,45] calculations were performed using SIESTA[46,47] program within generalized gradient approximation with Perdew, Bruke and Ernazerholf (PBE) functional for geometrical optimization through total energy calculation[48]. Norm-conserved Troullier-Martins pseudopotentials were used to describe the interaction between core and valence electrons[49]. The mesh cut off was 250 eV and Brillouin zone integration was sampled by 1x24x16 k-points within Monkhorst-pack scheme with double-ζpolarized basis set[50]. To simulate the unit cell of monolayers periodic boundary conditions were used with 24 Å vacuum space to minimize the interaction between the layers. The systems were fully relaxed to obtain the ground state structure with residual forces on the atoms less than 0.01 eV/atom. The electronic transport properties have been studied using non-equilibrium Green's function (NEGF) in TranSIESTA[51] module using 1x4x1 k-points grid which yields reasonably converged results. The transmission spectrum which defines the probability for electrons to be transferred from left to right electrode with the specific energy E, is calculated from the equation[52],

$$T(E,V) = \mathbf{tr}[\Gamma_R(E,V)G_C(E,V)\Gamma_L(E,V)G_C^+(E,V)[$$

Where $G_C$ is the Green's function of the central region and $\Gamma_{L/R}$ is the coupling matrix of electrodes in either sides. The integration of this transmission function gives the electric current,

$$I(V) = \int_{\mu_L}^{\mu_R} T(E,V(\{f(E-\mu_L) - f(E-\mu_R)\}dE$$

Where $\mu_L$= -V/2 ($\mu_R$=V/2) is the chemical potential of the left and right electrode.

## III. Results and Discussion:

### A. Structural and Electronic properties:

Structural and electronic properties of the two polymorphs are shown in Figure 1(a,b). $\beta_{12}$-borophene has 5 atom unit cell with *p2mm* symmetry and relaxed lattice parameter of a=2.92 and b=5.09 Å, which is in good agreement with already published results[21,28]. This is the one of the most stable structure among the 2D planar polymorphs of boron. Right panel in Figure 1(c) shows the band structure for $\beta_{12}$-borophene sheet, which clearly shows the metallic behavior and the bands near Fermi are mostly derived from $p_z$ orbital with some contribution from $p_y$ orbital. $\beta_{12}$-borophene induces some inhomogeneity as there are four fold (brown), fivefold (magenta) and six fold (purple) coordinated atoms in the unit cell. Unlike triangular borophene (diatomic basis) sheet, where the metallic bands were highly directional (present along only Γ-X and Γ-S directions)[53], here metallic bands are in all the directions. Y-direction of $\beta_{12}$-borophene structure is made of alternate boron linear chains. This gives rise to the metallic bands along Γ-Y, S-X and Γ-S where majority of them are from boron π-bond from $p_z$ orbital in linear chain and sigma bond from $p_x$, $p_y$ orbitals. In *x*-direction alternate filled hexagon pattern is found. Here, X-Γ bands come from $p_x$ and $p_y$ orbitals below Fermi, which give rise to σ-bond and above Fermi it shows contribution from $p_z$ orbital. Further, Y-S bands have contributions of $p_z$ orbital at Fermi level. For this polymorph, the lattice structure has inversion symmetry, where massless Dirac fermions are expected. These Dirac cones are above Fermi at an energy of ~ 2 eV and 0.5 eV. This upward shift in Dirac cones come from electron deficiency of boron and these positions can shift when the sheet will be placed in a metal substrate or strained as it can compensate the electron deficiency[29].

The second polymorph (Figure 1(b)), *χ*-borophene, is composed of 8 atoms in the unit cell (dotted lines), there are two type of boron atoms, 5-coordinated (purple) and 4-coordinated (brown). Lattice parameters are calculated to be a=2.94, b=8.50 Å, and it has metallic character showing in the band structure in right panel. Structural anisotropy in this structure is bigger than in $\beta_{12}$-borophene. This structure is characterized by chains of hollow sites separated by zigzag boron rows in *x*-direction and alternate chain of hollow sites shifted by half of the lattice in *y*-direction, which results planner *c2mm* symmetry. In the band structure as presented in Figure

1(d), shows metallic behavior of this polymorph. In this figure we can see that y-direction is dominated by boron $p_z$ orbitals along Γ-Y and X-S, and contribution from $p_y$ orbital is also found. There is gap in X-S direction, which gives rise to the anisotropy in the electronic structure. Similarly, bands in x-direction are composed mostly from σ-bonds. Along X-Γ bands, we have $p_x$ and $p_y$ character and along Y-S has weight of only $p_y$ orbital around the Fermi level. Γ-S has contributions from all the p-orbitals. This structure also possesses inversion symmetry responsible for the Dirac cone in this polymorph, in agreement with earlier reports[30]. Dirac cone is located ~ 0.4 eV below the Fermi level. These Dirac cones are anisotropic[29,30]. The observation of Dirac cones in both the above mentioned polymorphs, per se, is interesting and provides us greater scope to shift these along energy axis by several means (strain, substrate interaction etc.).

**B. Electron Transport Properties:**

A closer look into the intrinsic electronic transport properties help us to understand the suitability of a material for electronic device applications. Here, we present the results for electronic transport properties of both the borophenes considered in this work. The electronic transmission function is calculated by using semi-infinite leads (as left and right electrodes) and a device region as the central part. Right panel in Figure 2 shows the transport setup for these two polymorphs and left panel in Figure 2(a,b) shows the transmission function T(E) at zero bias for both the $β_{12}$-borophene and χ-borophene in x and y-directions. In Figure 2(a), $T_x(E)$, transmission of $β_{12}$-borophene in *x*-direction, has value of 0.80 channels/Å and $T_y(E)$, transmission in *y*-direction, has value of 0.79 channels/Å at the Fermi level. We see $T_x(E)$ has higher transmission above the Fermi level whereas $T_y(E)$ has higher values below the Fermi. This can be clearly explained from the band structure in Figure 1(c), where we have gap along Γ-Y and X-Γ has bands above the Fermi level, which lead to higher values for $T_x(E)$. Below the Fermi level Γ-Y has more bands than X-Γ, hence $T_y(E)$ has higher values. Transmission function from -1eV to 1eV does not show any effective anisotropy in the system.

Now we move to the case of χ-borophene. The anisotropy is higher when hollow sites are alternating in y-direction in this polymorph. Figure 2(b) shows that $T_x(E)$ has 1.2 channels/Å whereas $T_y(E)$ has 0.82 channels/Å at Fermi level. $T_x(E)$ has higher value than $T_y(E)$ beyond the

Fermi level as well. This can also be rationalized from the band structure in Figure1(d), where Γ-Y and X-S has only $p_z$ contribution around the Fermi whereas X-Γ and Y-S has contribution from both $p_x$ and $p_y$ around the Fermi level.

The bias is increased in a stepwise manner using NEGF and Landauer formula is used to calculate the current along x-direction (as $I_x$) and y-direction (as $I_y$). We applied bias up to 100 mV, see inset of Figure2(a). I-V characteristic shows negligible anisotropy in current for $\beta_{12}$-borophene. We calculated the current anisotropy in the $\beta_{12}$-borophene as η ($I_x/I_y$) =1.1 at 100 mV. The I-V characteristic for χ-borophene (the inset of Figure 2(b)) shows anisotropy in current and the calculated anisotropy comes out to be η ($I_x/I_y$) =1.5 at 100 mV. It is clear from the above discussion that structures belonging to the same family show different anisotropy in current and both of these have hollow sites in the form of hexagons. It is only the structural synergy which gives rise to the anisotropy. Anisotropy in these two polymorphs are smaller than the triangular borophene sheet reported earlier, where reported anisotropy was 2.1[53,54]. This anisotropy can, in principle, be exploited for practical device purposes and our study provides crucial insights in this direction.

## C. Effect of Strain on electronic and transport properties:

It is further established that the anisotropy can be tuned by using mechanical strain. So after successfully reproducing the structural and electronic properties of the pristine monolayer of these polymorphs, we now explore the effect of the strain on the electronic and transport properties. As these borophene sheets are unstable in free-standing form, so one will need a substrate to transfer the grown 2D sheet or to grow upon some substrate. Therefore, it is worthwhile to check the effect of strain on electronic and transport properties of these two sheets. We applied the uniaxial strain separately along the two mutually perpendicular directions, along x and y-direction. We first checked the effect of strain on the total energies of the system. Starting from the perfectly relaxed unit cell, strain ranging from 2% to 10% was applied. Applied strain is defined as $\varepsilon_x$= (a-$a_0$)/$a_0$ , $\varepsilon_y$= (b-$b_0$)/$b_0$, where a(b) and $a_0$($b_0$) are the lattice constants along the x(y) direction for the strained and relaxed structures, respectively. Strain of 2% would lead to deformation penalty in energy amounting to ~6-7 meV per boron atom in both structures, which is far below the thermal disturbance at room temperature. Earlier reports also suggest that imaginary phonons were found in pristine borophenes and these disappeared when 1% strain was applied[21]. Here we tried 2-10% of strain and we found that up to 6% of strain, the deformation energy was up to 60-70 meV/atom. Higher than 6% strain gives the deformation energy, which is greater than generally acceptable value of 150 meV/atom[55]. Electrical current can flow in the direction, not connected to the direction of the strain. We introduce the notation for transmission: in x-direction with strain in the same direction as $T_x$-*strainX* and with the strain in perpendicular direction as $T_x$-*strainY* same for transmission in y-direction with strain in perpendicular direction as *Ty-strainX* and transmission in y with strain in y-direction $T_y$-*strainY*.

**$β_{12}$-borophene:**

Figure 3(a-h) shows the band structures of strained $β_{12}$-borophene with 2 and 6% strains. First we would discuss the strain in x-direction which is in the upper panel in Figure 3(a-d). Tensile strain of 2% in x-direction opens gap close to Fermi at Γ-point, whereas Dirac cone is not affected as shown in Figure 3(b). The compressive strain closes the gap at Γ-point (Figure 3(c)). Effect of this gap change has already been explained in connection with the optical properties[55] . Band structure for 6% strain also shows that the tensile strain in x-direction further widens the gap at Γ-point up to 0.7 eV and Γ-S band shifted above the Fermi (Figure 3(a)). Band crossed along X-Γ and S-X bands become flat for compression in x-direction, where Γ-S band was pushed down the Fermi level as seen from Figure 3(d). We further move towards the strain in y-direction presented in Figure 3(e-h). 2% strain in y-direction also affects the band at Γ-point, but importantly it moves the Dirac cone in opposite direction for tensile and compressive strains shown in Figure 3(f,g). Compressive strain further opens the gap and tensile strain closes the gap at Γ-point. This is opposite of x-direction strain. Further, 6% compressive strain along y-direction also opens the gap at Γ-point and nearly closes it for tensile strain ( Figure 3(e-f)). Tensile strain gives almost flat band along S-X. It also moves the Dirac cone upwards upon tensile and downwards upon compressive strain.

Figure 4(a-d) shows the transmission function with 2% strain (and 6% strain insets) as $T_x$ and $T_y$ in x and y-directions respectively for $β_{12}$-borophene. Upper panel shows the transmissions in x-direction and lower panel shows the transmissions in y-direction. In the Figure 4(a) for *Tx-strainX*, we see that transmission states are getting shifted to lower or higher energy

side due to strains in either way; however, there is negligible change in transmission around Fermi. Above the Fermi level, compressive strain gives broad plateau in transmission than tensile strain. This happens due to the fact that tensile strain leads to gap along X-Γ and compressive strain closes it (Figure 3(b,c)). Below the Fermi level we have nearly same transmission for the both types of strains up to 0.6 eV; after that compressive strain gives higher transmission. Further increase of strain to 6% presented as insets of Figure 4(a), does not give significant changes in at the Fermi but beyond the Fermi level compressive strain gives higher transmission than tensile. This is because of a gap opening of 0.6 eV at Γ-point in tensile strained system while for compressive strain the bands are crossing the Fermi level (see in Figure 3(a,h)). Below the Fermi also, band along Y-S direction contributes for increase of transmission. Transmission in y-direction for strain in x-direction (*Ty-strainX*) (in Figure 4(b)), shows the changes around Fermi. It gives higher transmission when compressive strain is applied and low for tensile strain. Compressive strain closes the gap at Γ-point, which gives higher transmission close to Fermi. Above the Fermi level there is higher transmission for tensile strain and it is clear from Γ-Y bands. Below the Fermi compressive strain shows higher transmission; this is because of higher dispersion of band in Γ-Y (and Γ-S) directions, which effectively brings more states toward the Fermi level (Figure 3(b, c)). This happens due to linear atomic chain of boron atom in y-direction (Figure 1(a)). Strain in x-direction is perpendicular to boron atomic chain but it affects the bond length in the linear chain, which is responsible for the changes in transmission. Strain of 6% (inset of Figure 3(b)), yields lower transmission for both the tensile and compressive strain at the Fermi. Apart from this, tensile strain gives higher transmission above while compressive strain gives higher transmission below the Fermi level. It corresponds nicely with the changes in band structure in Γ-Y and Γ-S directions (see Figure 3(a,h)).

Now we shift our focus towards strain in y-direction. We see a broad band around Fermi for compressive strain while tensile strain gives a dip in x-direction transmission (*T$_x$-strainY*) at Fermi in Figure 4(c). This broad transmission plateau comes from X-Γ band crossing the Fermi level for compressive strain, while tensile strain opens a small gap at Fermi. Transmission above the Fermi level has broader plateau for tensile strain than the compressive one due to the bands along X-Γ. Below the Fermi, transmission is nearly same, but at 0.6 eV tensile strain gives higher transmission because of contributions from p$_y$ orbital along Y-S

(Figure 3(f, g)). Further increase in strain (inset Figure 4(c)), brings significant changes in transmission, namely compressive strain increases transmission at the Fermi level because $p_y$ band in X-Γ crosses the Fermi level. Above the Fermi, transmissions are nearly same for both the strains. Below the Fermi level, initially tensile strain demonstrates lower transmission but further below in energy transmission becomes higher than the compressive strain. It can be correlated to the changes in band structure in Y-S direction from Figure 3(a,h). Transmission in y-direction ($T_y$-strainY) as presented in Figure 4(d), have negligible changes for compressive strain but tensile strain shows a dip at Fermi. This dip is coming because of the gap at Γ point for the tensile strain while the band is crossing the Fermi for compressive strain at Γ-point in Figure 3(f, g). Linear stretch in boron atomic chain (Figure 1(a)) is responsible for this gap at Fermi level. Above the Fermi level, compressive strain shows higher transmission than tensile and below the Fermi they are nearly same. Inset of Figure 4(d) shows higher transmission for 6% compressive strain than tensile around the Fermi level and above the Fermi, and here the bands in Γ-Y direction are responsible; below the Fermi level transmissions are nearly same as shown in Figure 3(e,h). These changes in transmission/band structure happen because of linear chain of boron atoms in y-direction. As a consequence, the anisotropy in current can be tuned by application of strain, that we will discuss in the next section.

**χ-borophene:**

Now we move towards the χ-borophene, which is anisotropic in its pure form, as it has been discussed in previous sections. We can see the effect of strain on band structure of χ-borophene in Figure 5(a-h). 2% of strain can tune the gap at Γ-point, where $p_x$ and $p_y$ orbital are above Fermi and $p_z$ orbital is below Fermi at Γ-point. Tensile strain in x-direction (Figure 5(c)) pushed the $p_y$ orbital towards Fermi and $p_x$ orbital away from the Fermi, but compressive strain (Figure 5(b)) pushes $p_x$ orbitals down and $p_y$ away from Fermi. At Y-point, compressive strain pushes band to the Fermi level and tensile strain pushes it away from the Fermi. The rest of the bands are mostly unaffected upon strain. Y-direction strain affects the band at Γ-point inversely. Tensile strain (in Figure 5(g)) affects the $p_x$ and pushed it down towards Fermi and shifted $p_y$ away from it and makes the gap at Γ-point bigger. Compressive strain in y-direction shown in Figure 5(f), decreases the gap Γ-point and it shifts the $p_y$ orbital towards Fermi level and $p_x$ orbital away from Fermi. Further, strain of 6% affect the bands around the Fermi in the same way. Compressive strain in x-direction (Figure 5(a)), opens up the gap at Γ-point about 2.3 eV whereas closes it up to 0.6 eV for tensile strain as evident from Figure 5(d). The bands originated from $p_x$ and $p_y$ states at Γ-point come close for compressive and go far from each other for tensile strain. Band at Y-point comes close to Fermi in compressive strain and goes away in tensile strain. In y-direction, tensile strain (Figure 5(h)) opens up the gap at Γ-point around 2 eV and closes it for compressive strain up 1.2 eV (Figure 5(e)). Tensile strain pushes $p_x$ and $p_y$ orbital close to each other whereas for compressive strain they go far from each other. Band at Y-point also behave inversely to the strain in x-direction. Effect of strain on χ-borophene is less than $β_{12}$-borophene. Movement of Dirac cone is also smaller than that of $β_{12}$-borophene. This suggests that for χ-borophene transport properties would be robust with respect to strain compared to $β_{12}$-borophene.

Figure 6(a-d) shows the effect of 2% strain (inset 6% strain) in transmission for χ-borophene. We first discuss the 2% strain in x-direction. Tensile strain in x-direction (Figure 6(a)), does not affect the transmission $T_x$-*srainX* at Fermi level, the resonance like features should be attributed to the effect of bond stretching. Above the Fermi level, the transmission is suppressed for tensile strain because of the fact that Y-S bands go away from the Fermi level whereas for compressive strains they come close to it. Overall transmission above 0.5eV is high for compressive strain due to band structure change in X-Γ direction (Figure 5(b, c)). Further 6% strain in x-direction gives no change at Fermi, but overall transmission above the Fermi level is higher for compressive strain than the tensile one (inset of Figure 6(a)). It is due to band related to $p_x$ orbital, that comes closer to Fermi for compressive strain and goes away for tensile strain. Y-S band also comes closer to Fermi. Below the Fermi level, initially tensile strain shows higher transmission but compressive strain overcomes this at 0.2 eV. This is because of a band rising in energy in X-Γ direction in Figure 5(a,d). $T_y$-*strainX* in Figure 6(b), does not change at Fermi level but above the Fermi at 0.5 eV, tensile strain has higher transmission than compressive strain because of $p_y$ orbital changes in Γ-Y direction. Below the Fermi level the transmission is nearly the same (Figure 5(b, c)). Inset of Figure 6(b) shows that tensile strain has higher transmission than compressive one at the Fermi because of S-X band. Overall transmission above Fermi is also high for tensile strain because of a gap at Γ-point. Below the Fermi, transmission is high for compressive strain because of S-X band (Figure 5(a,d)).

Strain in y-direction gives the change in $T_x$-strainY (Figure 6(c)), where transmission is getting higher for compressive strain and lower for tensile strain. But these changes are not significant at the Fermi level. Above Fermi, tensile strain gives us the broad band in transmission. This is due to bands at Y-S come close to Fermi level in Figure 5(f, g). If we further see above 0.4eV, compressive strain gives higher transmission because the gap at Γ-point got reduced. Below the Fermi level, over all transmission for compressive strain is higher. Further 6% strain shows higher transmission for compressive strain than the tensile strain at the Fermi (inset of Figure 6(c)). It is due gap at the Γ-point that is lower for compressive than tensile strain. Just above the Fermi, tensile strain starts giving higher transmission, this is because of Y-S band. Below the Fermi, initially compressive strain gives higher transmission but at -0.7 eV tensile starts giving higher transmission is because of X-Γ band (Figure (e,h)). $T_y$-strainY in Figure 6(d), does not show significant changes upon strain around Fermi. Above the Fermi compressive strain has broad transmission because of the gap at Γ-point, while for tensile strain it behaves inversely as seen in Figure 5(f, g). Below the Fermi, transmission is nearly the same. Inset of Figure 6(d), shows that over all transmission is high above Fermi level for compressive strain than tensile strain. It is due to a band which goes away from Fermi for tensile strain in Γ-Y direction (Figure 5(e,h)). Below the Fermi, transmission is nearly same and at -0.8 eV compressive strain gives higher transmission.

**D. I-V characteristics:**

We have calculated the I-V characteristics for both the polymorphs for the low bias range up to 100 mV to see how does strain affects the anisotropy and current. Figure 7(a-d) shows the I-V characteristics for all the directions and strains in $β_{12}$-borophene. $I_x$-strainX remains unchanged upon both tensile and compressive strain (Figure 7(a)). For 6% of strain, $I_x$-strainX manifest quite same current at 100 mV. On the other hand, in Figure 7(b), $I_y$-strainX changes ~10% upon tensile and ~15% upon compressive strain at 100 mV and increment of strain gives less current for both the tensile and compressive strains. Strain in y-direction gives significant changes in current. We calculated changes in current upon 2% strain to be ~9% for tensile and ~22% for compressive strain in $I_x$-strainY. For the case of 6% strain, we see 26% increment for compressive and 31% decrement for tensile strain (Figure 7(c)). $I_y$-strainY in Figure 7(d), leads to a change of 17% for compressive and 12% in tensile strains of 2%. Further increment in strain gives the highest deviation in current of 34% for both tensile and compressive strain. We also calculated the anisotropy up to η ($I_x/I_y$)=1.34-1.64 at 100 mV maximum for 2-6% tensile strain in *x*-direction whereas compressive strain in *x*-direction is quietly unaffected. Rest of the strains does not affect the anisotropy much. We found y-direction strain offers higher change in current values but x-direction strain increases anisotropy. It is quite clear that strain in y-direction changes the electronic properties of boron atomic chain, which influences the transmission and current. Whereas in x-direction, application of strain gets compensated by filled hexagons (see Figure1(a)).

The changes in I-V characteristics for χ-borophene as shown in Figure 7(e-h) are not as significant as that of $β_{12}$-borophene. Maximum change happens along the strain in y direction, where *$I_y$-strainY* increases to 3.7% on compressive and decreases to 5% on tensile strain. Further, 6% strain gives maximum 12% changes for both the compressive and tensile see in Figure 7(h). These changes are way lower in comparison to the change in current in $β_{12}$-borophene. Apart from this, we do not see much change in current in other directions due to strain. Anisotropy is preserved around η($I_x/I_y$) =1.5 at 100 mV for 2% of strain in both the directions. Even after 6% of strain, anisotropy is preserved and we get maximum in current ratio as η($I_x/I_y$)=1.54 at 100 mV. It shows that χ-borophene is robust in the terms of electronic transport with the applied strain. This is happening because of the structural properties of χ-borophene. In χ-borophene, hollow hexagons are not in line, they are alternated in structure (see figure 1(b)), which compensates the effect of strain in the terms of bond stretch. There is no linear atomic chain of boron atoms in χ-borophene, which also reduces the effect of strain in the structure. We further conclude from these I-V characteristics that the $β_{12}$-borophene has tunable transport properties and χ-borophene is robust in nature.

## IV. Conclusions:

Using density functional theory (DFT) combined with NEGF approach, we have investigated electronic and transport properties of the two recently synthesized 2D boron polymorphs namely $β_{12}$-borophene and χ-borophene. To begin with, we have focused on the pristine systems, where $β_{12}$-borophene shows isotropic and χ-borophene shows anisotropic transport properties with

$\eta(I_x/I_y)=1.5$, which is supported by their underlying electronic structures. When these polymorphs are subjected to the external unidirectional strain, in addition to electronic structure modification we also see the tunable anisotropy in the IV characteristics, particularly in $\beta_{12}$-borophene. Furthermore, $\chi$-borophene remains robust and nearly unaffected by the external strain. Our calculation reveals that the anisotropy in current ($\eta=I_x/I_y$) in $\beta_{12}$-borophene can be tuned from 1.1 to 1.64 upon 6% of strain in x-direction. This significant tuning of anisotropy is an interesting finding where the compensation for strain energy is only limited below 100 meV/atom. Keeping in hindsight that the 2D borophene structures have limited stability in their free standing forms, any realistic approach to grow monolayers of borophene will amount to some degree of strain in the structure. Further, keeping our results alongside this perspective, it seems that 2D borophenes are very intriguing and hold tremendous possibilities for applications in electronic devices.

**Acknowledgements:**

The authors acknowledge computational resources provided through Swedish National Infrastructure for Computing (SNIC2017-11-28, SNIC2017-5-8 SNIC2017-1-237). VS acknowledges funding from the European Erasmus fellowship program. AG and RA acknowledge support from the Swedish Research Council.

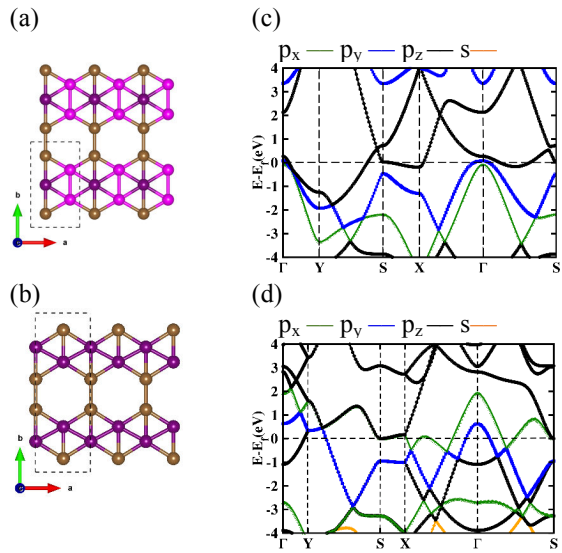

Figure 1. Atomic structures of (a) $\beta_{12}$-borophene and (b) $\chi$-borophene (dotted lines represent the unit cell), and band structures for (c) $\beta_{12}$-borophene and (d) $\chi$-borophene

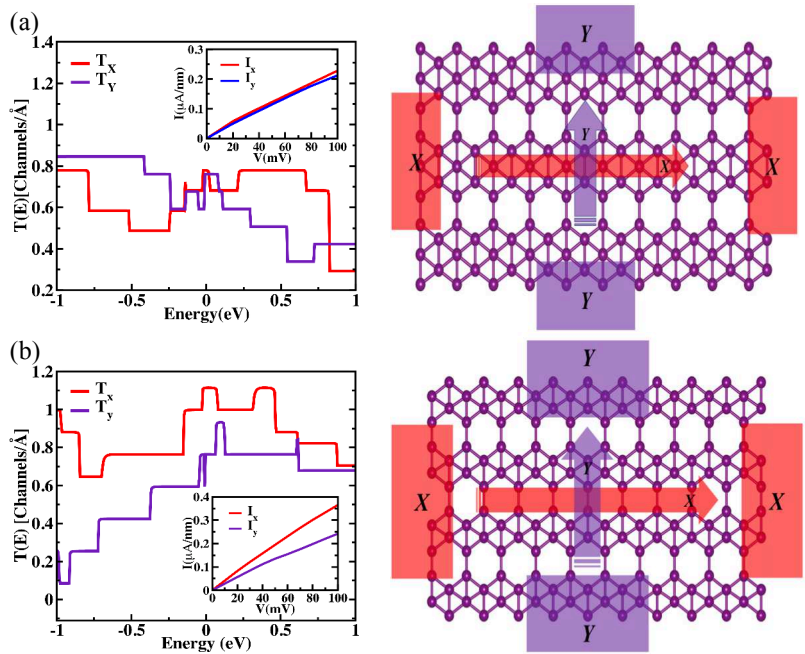

Figure 2. Schematic representation of transport set up (right panel) and zero bias transmission functions (left panel) along x ($T_x$) and y ($T_y$) directions with the corresponding I-V characteristics in x ($I_x$) and y ($I_y$) directions (inset figures) for (a) $\beta_{12}$-borophene and (b) for $\chi$-borophene

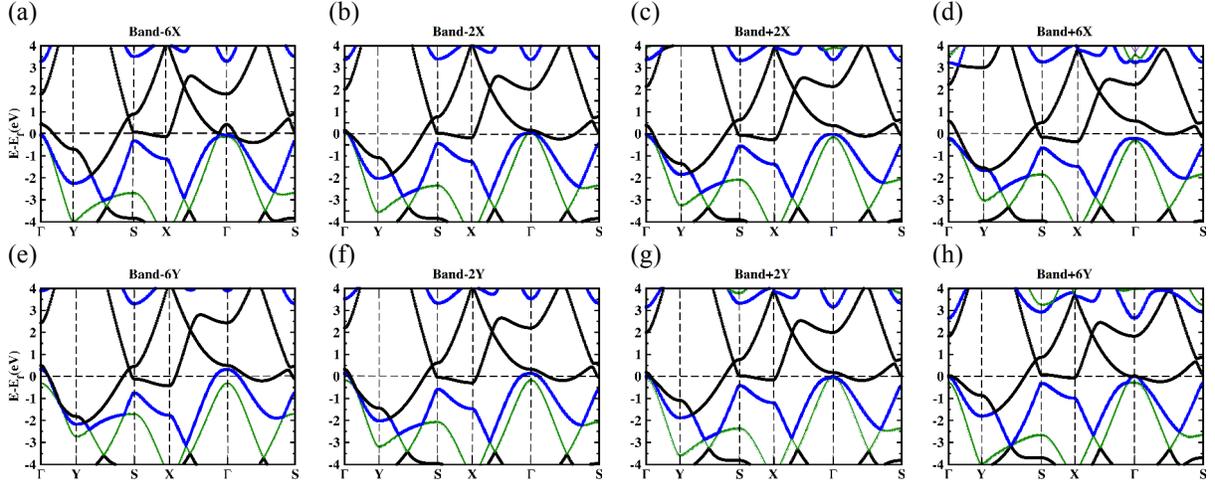

Figure 3. Band structure for unidirectional compressive and tensile strain along x-direction (upper panel, (a)-(d)) and along y-direction (lower panel, (e)-(h)) in $\beta_{12}$-borophene. (The % of applied strain is given in the figure titles where + denotes tensile strain and – as compressive strain)

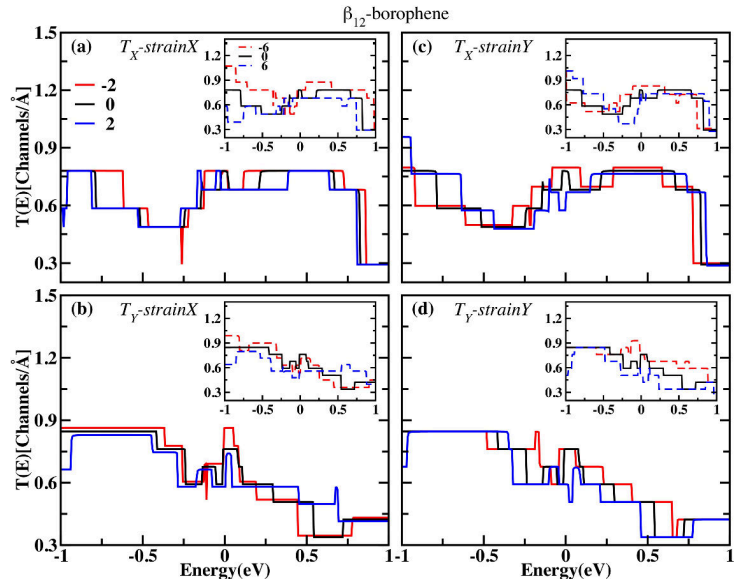

Figure 4. (a)-(d) Zero bias transmissions ($T_x$) and y ($T_y$) for $\beta_{12}$-borophene with applications of strain along x and y directions. The main panels show results for 2% of strain and the insets correspond to 6% of strain.

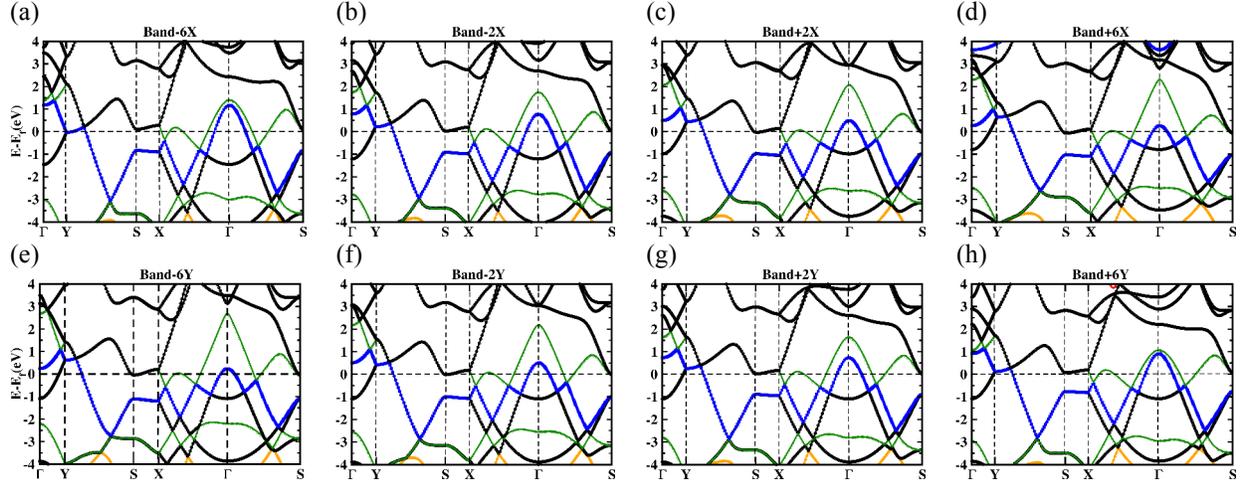

Figure 5. Band structure for compressive and tensile strain along *x*-direction (upper panel, (a)-(d)) and along *y*-direction (lower panel, (e)-(f)) of $\chi$-borophene (The % of applied strain is given in the figure titles where +denotes tensile strain and – as compressive strain)

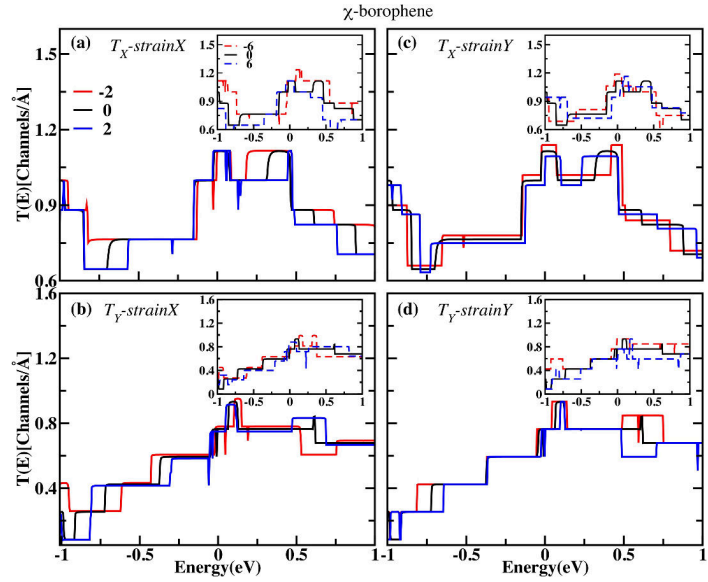

Figure 6. (a)-(d) Zero bias transmissions ($T_x$) and y ($T_y$) for $\chi$-borophene with applications of strain along x and y directions. The main panels show results for 2% of strain and the insets correspond to 6% of strain.

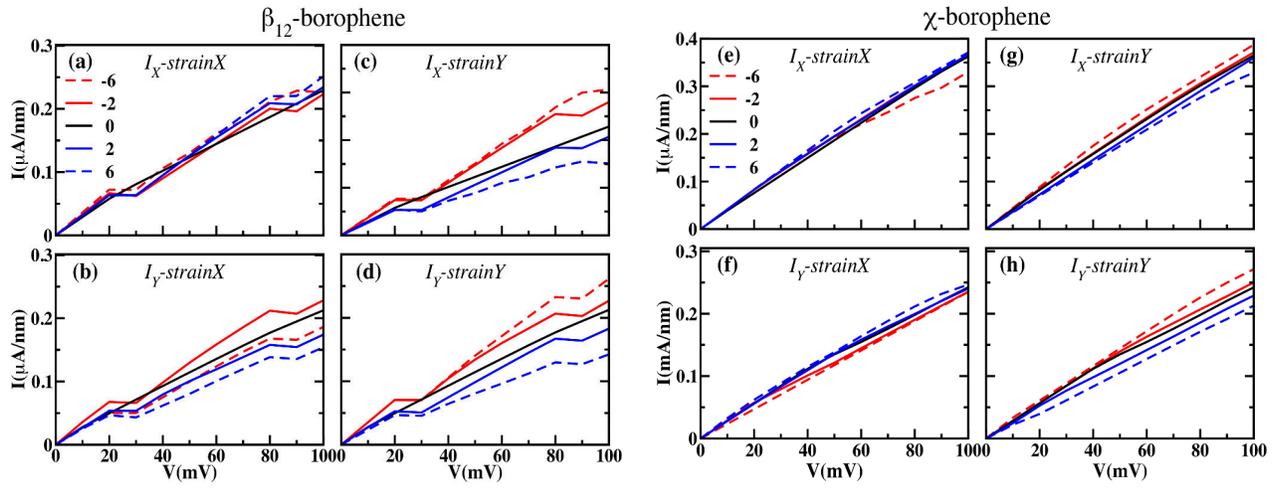

Figure 7. (a)-(d) I-V characteristics for $\beta_{12}$-borophene, (e)-(h) $\chi$-borophene along x and y directions with application of 2% and 6% unidirectional strain along x and y.